\documentclass[twoside,fleqn]{ActaStyle}
\usepackage{amsmath}
\usepackage{epsfig}
\def\authorheading{Holger Gies, Christof Wetterich}
\def\shorttitle{Renormalization flow from UV to IR degrees of freedom}

\newcommand{\lNJL}{\lambda_{\text{NJL}}}
\newcommand{\xsb}{$\chi$SB}
\newcommand{\I}{\text{i}}

\newcommand{\re}[1]{~(\ref{#1})}

\newcommand{\Gk}{\Gamma_k}
\newcommand{\yl}{\psi_{\text{L}}}
\newcommand{\yr}{\psi_{\text{R}}}
\newcommand{\ybl}{\bar{\psi}_{\text{L}}}
\newcommand{\ybr}{\bar{\psi}_{\text{R}}}
\newcommand{\yb}{\bar{\psi}}

\newcommand{\lk}{\bar{\lambda}_{\sigma,k}}
\newcommand{\mkq}{\bar{m}_k^2}
\newcommand{\hk}{\bar{h}_k}

\newcommand{\dlk}{\Delta\lk}

\newcommand{\fss}[1]{#1\!\!\!/}
\newcommand{\fsl}[1]{#1\!\!\!\!/}
\newcommand{\SmP}{\ybr\yl\ybl\yr}
\newcommand{\Yc}{(\ybr\yl \phi-\ybl\yr\phi^\ast)}
\newcommand{\phid}{\phi^\ast}

\newcommand{\pat}{\partial_t}

\newcommand{\kB}{\Lambda}
\newcommand{\te}{\tilde{\epsilon}}

\newcommand{\Nc}{N_{\text{c}}}
\begin{document}
\pagerange{1}{7}

\title{RENORMALIZATION FLOW\\ FROM UV TO IR DEGREES OF FREEDOM\footnote{Talk given by H.G.~ at the conference RG-2002, March 10-16, 2002, Strba, Slovakia.}} 
\author{Holger Gies\email{holger.gies@cern.ch}}{TH Division, CERN, CH-1211 Geneva 23, Switzerland}
\author{Christof Wetterich\email{C.Wetterich@thphys.uni-heidelberg.de}}{Institut f\"ur theoretische Physik,
  Universit\"at Heidelberg,\\ Philosophenweg 16, D-69120 Heidelberg,
  Germany} 

\day{April 18, 2002}

\abstract{Within the framework of exact renormalization group flow
  equations, a scale-depen\-dent transformation of the field variables
  provides for a continuous translation of UV to IR degrees of
  freedom. Using the gauged NJL model as an example, this translation
  results in a construction of partial bosonization at all scales.  A
  fixed-point structure arises which makes it possible to distinguish
  between fundamental-particle and bound-state behavior of the scalar
  fields.}


\vspace{-11cm}
\hspace{7cm} CERN-TH/2002-105

\vspace{11cm}

\section{Introduction}
For an investigation of a system of interacting quantum fields, it is
mandatory to identify the ``true'' degrees of freedom of the system.
As we know from many physical systems such as QCD, or the plethora of
condensed-matter systems, the nature of these degrees of freedom of
one and the same system can be very different at different momentum
(or length) scales.  Of course, the first physical task is the
identification of these relevant degrees of freedom at the various
scales.  Simplicity can be an appropriate criterion for this, in
particular, simplicity of the effective action governing these degrees
of freedom.

Whereas quantum field theory is usually defined in terms of a
functional integral over quantum fluctuations of those field variables
that correspond to the degrees of freedom in the ultraviolet (UV), we
are often interested in the properties of the system in the infrared
(IR). In some but rare instances, we know not only the true degrees of
freedom at these different scales, but also the formal translation
prescription of one set of variables into the other in terms of a
discrete integral transformation. An example is given by the
Nambu--Jona-Lasinio (NJL) model \cite{E} in which self-interacting
fermions (UV variables; ``quarks'') can be translated into an
equivalent system of (pseudo-)scalar bosons (IR variables; ``mesons'')
with Yukawa couplings to the fermions. This is done by means of a
Hubbard-Stratonovich transformation, also called partial
bosonization. A purely bosonic theory can then be obtained by
integrating out the fermions.

Integrating out the fermions at once leads, however, to highly
nonlocal effective bosonic interactions. This problem can be avoided
by integrating out the short distance fluctuations stepwise by means
of the renormalization group. In this context, a continuous
translation from multifermion to bosonic interactions would be
physically more appealing, since it would reflect the continuous
transition from the ultraviolet to the infrared more naturally.
Furthermore, phases in which different degrees of freedom coexist
could be described more accurately.

In the following, we will report on a new approach which is capable of
describing such a continuous translation. The approach is based on an
exact renormalization group flow equation for the effective average
action \cite{Wetterich:1993yh} allowing for a scale-dependent
transformation of the field variables \cite{Gies:2002nw}.\footnote{For
  an earlier approach, see \cite{Ellwanger:1994wy}. A general account
  of field transformations within flow equations has been given in
  \cite{Latorre:2000qc}.} In order to keep this short presentation as
transparent as possible, we will discuss our approach by way of
example, focussing on the gauged version of the NJL model which shares
many similarities with, e.g., building blocks of the standard model.

We shall consider the gauged NJL model for one fermion flavor in its
simplest version characterized by two couplings in the UV: the gauge
coupling $e$ of the fermions to an abelian gauge field $\sim
\yb\fsl{A}\,\psi$, and the chirally invariant four-fermion
self-interaction in the (pseudo-)scalar channel $\sim\SmP$ with
coupling $\lNJL$. Depending on the values of these couplings, the
gauged NJL model interpolates between the pure NJL model entailing
chiral-symmetry breaking (\xsb) for strong $\lNJL$ coupling and
(massless) QED for weak $\lNJL$; for simplicity, the gauge coupling is
always assumed to be weak in the present work. The physical properties
and corresponding degrees of freedom in the infrared depend crucially
on $\lNJL$: we expect fermion condensates and bosonic excitations on
top of the condensate in the case of strong coupling, but bound states
such as positronium at weak coupling. We shall demonstrate that our
flow equation describes these features in a unified manner. The
question as to whether the fields behave like fundamental particles or
bound states thereby receives a scale-dependent answer; in particular,
this behavior can be related to a new infrared fixed-point structure
with interesting physical implications.

\section{Fundamental particles versus bound states}
Let us study the scale-dependent effective action $\Gk$ for the abelian
gauged NJL model ($N_{\text{f}}=1$) including the scalars arising from
bosonization in the following simple truncation,
\begin{eqnarray}
\Gk&=&\int d^4x\biggl\{ \yb\I\fss{\partial}\psi +2\lk\,\SmP
-e\yb\fsl{A}\,\psi+\frac{1}{4} F_{\mu\nu}F_{\mu\nu} \nonumber\\
&&\qquad\qquad+Z_{\phi,k}\partial_\mu\phid\partial_\mu\phi+\mkq\,
\phid\phi +\hk\Yc  \biggr\}
  \nonumber, \label{1}
\end{eqnarray}
where we take over the conventions from \cite{Gies:2002nw}. Beyond the
kinetic terms, we focus on the fermion self-interaction $\sim \lk$,
the scalar mass $\sim \mkq$, and the Yukawa coupling between the
fermions and the scalars $\sim \hk$. In the framework of exact RG
equations, the infrared scale $k$ divides the quantum fluctuations
into modes with momenta $k<p<\Lambda$ that have been integrated out,
so that $\Gk$ governs the dynamics of those modes with momenta $p<k$
which still have to be integrated out in order to arrive at the full
quantum effective action $\Gamma_{k\to 0}$. The RG flow of $\Gamma_k$
to the quantum effective action is described by a functional
differential equation \cite{Wetterich:1993yh} which we solve within
the truncation given by Eq.\re{1}. The flow is initiated at the UV
cutoff $\Lambda$, which in our case also serves as the bosonization
scale, and we fix the couplings according to
\begin{equation}
\lambda_{\text{NJL}}=\frac{1}{2}\,
  \frac{\bar{h}_{\Lambda}^2}{\bar{m}_{\Lambda}^2}, \quad
  \bar{\lambda}_{\sigma,\Lambda}=0, \quad Z_{\phi,\Lambda}=0
\label{1.6}. 
\end{equation}
In other words, all fermion self-interactions are put into the Yukawa
interaction $\hk$ and the scalar mass $\mkq$ at the bosonization scale
$\Lambda$, and the standard form of the gauged NJL model in a purely
fermionic language could be recovered by performing the Gaussian
integration over the scalar field.

Concentrating on the flow of the couplings $\mkq,\hk,\lk$, we
find\footnote{The numerical coefficients on the RHS's of Eqs.\re{23}
  depend on the implementation of the IR cutoff procedure at the scale
  $k$ and on the choice of the Fierz decomposition of the four-fermion
  interactions. For the former point, we use a linear cutoff
  function \cite{Litim:2001up} (see also D.F.~Litim's contribution to
  this volume). For the latter, we choose a $(S-P)$, $(V)$
  decomposition, but display only the (pseudo-)scalar channels here;
  the vectors are discussed in \cite{Gies:2002nw}. Furthermore, we
  work in the Feynman gauge.} ($\pat\equiv k (d/dk)$):
\begin{eqnarray}
\pat \mkq&=& \frac{k^2}{8\pi^2}\, \hk^2,\nonumber\\
\pat\hk&=& -\frac{1}{2\pi^2}\,e^2\,\hk+{\cal O}(\lk),\label{23}\\
\pat\lk&=&-\frac{9}{8\pi^2 k^2}\, e^4
  +{\frac{1}{32\pi^2 Z_{\phi,k}^2 k^2}\,
  { \frac{3 + \frac{\mkq}{Z_{\phi,k}
          k^2}}{(1 + \frac{\mkq}{Z_{\phi,k} k^2})^3}}}\,
\hk^4+{\cal O}(\lk).\nonumber
\end{eqnarray} 
We observe that, although the four-fermion interaction has been
bosonized to zero at $\Lambda$, $\bar{\lambda}_{\sigma,\Lambda}=0$,
integrating out quantum fluctuations reintroduces four-fermion
interactions again owing to the RHS of the last equation; for
instance, the first term $\sim e^4$ arises from gauge boson exchange.
Bosonization in the standard approach is obviously complete only at
$\Lambda$. However, guided by the demand for simplicity of the
effective action at any scale $k$, we would like to get rid of the
fermion self-interaction at all scales, i.e., re-bosonize under the
flow. Here the idea is to employ a flow equation for a scale-dependent
effective action $\Gamma_k[\phi_k]$, where the scalar field variable
$\phi_k$ is allowed to vary during the flow; this flow equation is
derived in \cite{Gies:2002nw}, and can be written in a simple form as
\begin{equation}
\pat\Gamma_k[\phi_k]=
  \pat\Gamma_k[\phi_k]\bigl|_{\phi_k}
  +\int_q\left( \frac{\delta\Gamma_k}{\delta\phi_k(q)}\, \pat\phi_k
  (q) + \frac{\delta\Gamma_k}{\delta\phid_k(q)}\, \pat\phid_k(q)\right),
  \label{ffv16} 
\end{equation}
where the notation omits the remaining fermion and gauge fields for
simplicity. The first term on the RHS is nothing but the flow equation
for fixed variables evaluated at fixed $\phi_k$ instead of
$\phi=\phi_\Lambda$.  The second term reflects the flow of the
variables. In the present case, we may choose
\begin{equation}
\pat\phi_k(q)=-(\ybl\yr)(q)\, \pat\alpha_k, \quad
\pat\phid_k(q)=(\ybr\yl)(-q)\, \pat\alpha_k,\label{d13b}
\end{equation}
where the transformation parameter $\alpha_k(q)$ is an a 
priori arbitrary function. Projecting
Eq.\re{ffv16} onto our truncation\re{1}, we arrive at modified flows
for the couplings (the equation for $\mkq$ remains unmodified): 
\begin{eqnarray}
\pat\hk&=&\pat\hk\bigl|_{\phi_k}+\mkq\,
  \pat\alpha_k, \label{ffv17}\\
\pat\lk&=&\pat\lk\bigl|_{\phi_k}-\,\hk\,
\pat \alpha_k. \nonumber
\end{eqnarray}
We can now obtain bosonization at all scales, $\lk=0$, if we adjust
$\alpha_k$ in such a way that the RHS of the $\pat\lk$ equation equals
zero for all $k$. This, of course, affects the flow of the Yukawa
coupling $\hk$. The physical effect can best be elucidated with the
aid of the convenient coupling $\te_k:=\frac{\mkq}{k^2\hk^2}$ and its
RG flow:
\begin{equation}
\pat \te_k=-2\te_k + \frac{1}{8\pi^2} 
  +\frac{e^2}{\pi^2} \te_k + \frac{9e^4}{4\pi^2} \te_k^2
  -\frac{1}{16\pi^2}\frac{\epsilon_k^2(3+\epsilon_k)}{(1+\epsilon_k)^3},
  \label{C} 
\end{equation}
where we also abbreviated $\epsilon_k:=\frac{\mkq}{Z_{\phi,k}k^2}$. A
schematic plot of $\pat\te_k$ is displayed in Fig.~1 where the
occurence of two fixed points is visible (note that all qualitative
features discussed here are insensitive to the last term of
Eq.\re{C}). The first fixed point $\te^\ast_1$ is infrared unstable
and corresponds to the inverse critical $\lNJL$ coupling. Starting
with an initial value of $0<{\te}_\Lambda<\te^\ast_1$ (strong
coupling), the flow of the scalar mass-to-Yukawa-coupling ratio will
be dominated by the first two terms in the flow equation\re{C} $\sim
-2 \te_k+1/(8\pi^2)$. This is a typical flow of a theory involving a
``fundamental'' scalar with Yukawa coupling to a fermion sector.
Moreover, we will end in a phase with (dynamical) chiral symmetry
breaking, since $\tilde\epsilon\sim \mkq$ is driven to negative
values.

\begin{figure}[t]
\begin{picture}(120,30)
\put(30,0){
\epsfig{figure=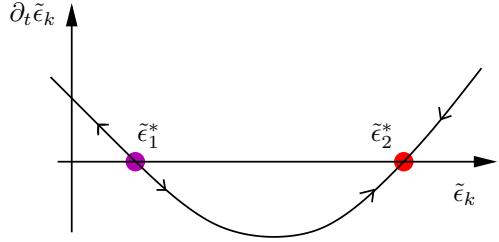,width=6cm}}
\put(26,29){$\pat\te_k$}
\put(43,13){$\te_1^\ast$}
\put(74,13){$\te_2^\ast$}
\put(85,5){$\te_k$}
\end{picture} 

\caption{Schematic plot of the fixed-point structure of the $\te_k$
  flow equation after fermion-boson translation. Arrows indicate the
  flow towards the infrared, $k\to 0$.}
\end{figure}

On the other hand, if we start with $\te_\Lambda>\te^\ast_1$, the flow
will necessarily be attracted towards the second infrared-stable fixed
point $\te^\ast_2$. There will be no dynamical symmetry breaking,
since the mass remains positive.  The effective four-fermion
interaction corresponding to the second fixed point reads:
$\lambda^*_\sigma=\frac{1}{2k^2\tilde\epsilon^*_2}\approx
\frac{9}{16\pi^2} \frac{e^4}{k^2}$, which coincides with the
perturbative value of massless QED. We conclude that the second fixed
point characterizes massless QED. The scalar field shows a typical
bound-state behavior with mass and couplings expressed by $e$ and
$k$. A more detailed analysis reveals that the scalar field
corresponds to positronium at this fixed point \cite{Gies:2002nw}. 

Our interpretation is that the ``range of relevance'' of these two
fixed points tells us whether the scalar appears as a ``fundamental''
or a ``composite'' particle, corresponding to the state of the system
being governed by $\te^\ast_1$ or $\te^\ast_2$, respectively.

\section{Physics at the bound-state fixed point}
The bound-state fixed point $\te^\ast_2$ shows further interesting
physical properties. In order to unveil them, we have to include
momentum dependences of the couplings; in particular, we study the
momentum dependence of $\lk$ in the $s$ channel. Then we can
generalize the fermion-to-boson translation\re{d13b},
\begin{equation}
\pat\phi_k(q)=-(\ybl\yr)(q)\, \pat\alpha_k(q)+ \phi_k(q)\,
\pat\beta_k(q),\label{d13ba}
\end{equation}
(and similarly for $\phid$) with another a priori arbitrary function
$\beta_k(q)$. Now we can fix $\alpha_k(q)$ and $\beta_k(q)$ in such a
way that $\lk(s)$ vanishes simultaneously for all $s$ and $k$ and that
$\hk$ becomes momentum-independent. Defining the dimensionless
renormalized couplings $\epsilon_k={\mkq}/({Z_{\phi,k}k^2}), h_k=\hk
\, Z_{\phi,k}^{-1/2}$, this procedure leads us to the final flow
equations \cite{Gies:2002nw},
\begin{eqnarray} 
\pat \epsilon_k&=& -2 \epsilon_k + \frac{h_k^2}{8\pi^2}
-\frac{\epsilon_k(\epsilon_k+1)}{h_k^2} \left(\frac{9e^4}{4\pi^2}
  -\frac{h_k^4}{16\pi^2} \frac{3+\epsilon_k}{(1+\epsilon_k)^3} \right)
\bigl(1+(1+\epsilon_k)Q_\sigma\bigr), \nonumber\\
\pat h_k&=& -\frac{e^2}{2\pi^2}\, h_k
-\frac{2\epsilon_k+1+(1+\epsilon_k)^2Q_\sigma}{h_k}\left(
  \frac{9e^4}{8\pi^2} -\frac{h_k^4}{32\pi^2}
  \frac{3+\epsilon_k}{(1+\epsilon_k)^3} \right). \label{B}  
\end{eqnarray}
Using $\te_k=\epsilon_k/h_k^2$, Eq.\re{C} can be rediscovered from
Eqs.\re{B}. Defining $\dlk:=\lk(k^2)-\lk(0)$, the quantity
$Q_\sigma\equiv \pat\dlk/\pat\lk(0)$ measures the suppression of
$\lk(s)$ for large external momenta. Without an explicit computation,
we may conclude that this suppression implies $Q_\sigma<0$ in
agreement with unitarity (e.g., $Q_\sigma\simeq -0.1$). In Fig.~2, a
numerical solution of Eqs.\re{B} is presented in which we release the
system at $\Lambda$ at $\te_\Lambda>\te_1^\ast$, so that it approaches
$\te_2^\ast$ in the IR.

\begin{figure}[t]
\begin{picture}(120,35)
\put(30,0){
\epsfig{figure=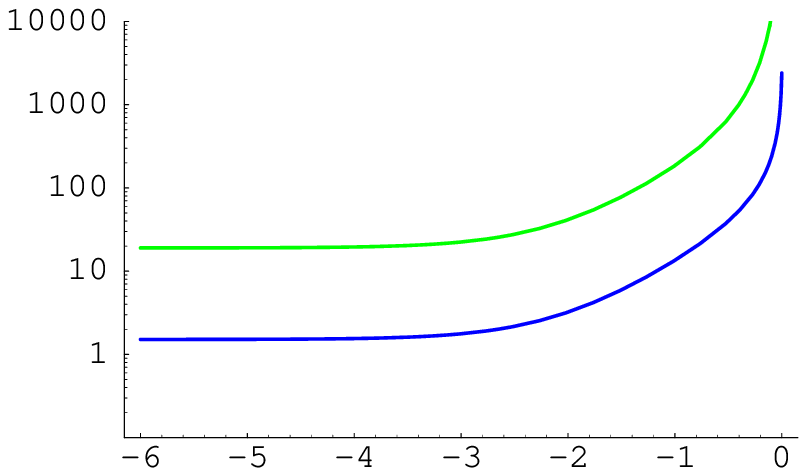,width=6cm}}
\put(83,31){$\epsilon_k$}
\put(90,21){$h_k$}
\put(85,0){$t$}
\put(30,3){IR}
\put(100,3){UV}
\end{picture} 

\caption{Flows of $\epsilon_k$, $h_k$ in the
  symmetric phase according to Eqs.\re{B}
  for the initial values $\epsilon_{\kB}=10^6$,
  $\te_\Lambda=0.17>\te_1^\ast$, $e=1$, $Q_\sigma=-0.1$ ($t=\ln
  k/\Lambda$).} 
\end{figure}
 
We observe that both $h_k$ and $\epsilon_k$ approach fixed points in
the IR. (For analytical results for the fixed points, see
\cite{Gies:2002nw}.) In particular, this implies that the scalar mass
term $m_k^2=\epsilon_k k^2$ decreases with $k^2$ in the symmetric
phase. This is clearly a nonstandard running of a scalar particle
mass. As a consequence, a large scale separation $\Lambda\gg k$ gives
rise to a large mass scale separation $m_\Lambda\gg m_k$ without any
fine-tuning of the initial parameters.

\section{Conclusions and Outlook}
Within the framework of exact renormalization group flow equations for
the effective average action, a scale-dependent transformation of the
field variables provides for a continuous translation of UV to IR
degrees of freedom. This concept is able to realize the physical
criterion of desired simplicity of the effective action. Using the
gauged NJL model as an example, this translation can be regarded as
partial bosonization at all scales. Here we identified an infrared
fixed-point structure which can be associated with a bound-state
behavior. One main result is that the RG flow of the scalar mass at
the bound-state fixed point is ``natural'' in 't Hooft's sense so that
no fine-tuning problem arises if we want to have small masses at
scales far below the UV cutoff $k\ll \Lambda$.

It should be interesting to see if this possibility of a naturally
small scalar mass is applicable for the gauge hierarchy problem
of the standard model. For this purpose, a mechanism has to be
identified that causes the system to flow into the phase with
spontaneous symmetry breaking after it has spent some ``RG time'' at
the bound-state fixed point. Phrased differently, the bound-state
fixed point has to disappear in the deep IR.  Taking a first glance at
Eq.\re{C}, or its immediate nonabelian generalization for
SU($N_{\text{c}}$) gauge groups (here we use the Landau gauge),
\begin{equation}
\pat\te_k
=-\left(2-\frac{3\,{C_2} }{4\pi^2}\,{g_k}^2\right)
  \te+\frac{{\Nc}}{8\pi^2}  
  +\frac{9}{8\pi^2}\frac{{C_2}}{{\Nc}} \left({C_2}
       -\frac{1}{2{\Nc}}\right)  \, {g_k}^4\, \te^2 +{ \cal
       O}(\epsilon^2), 
\label{nonab}
\end{equation}
where $g_k$ is the running gauge coupling and
$C_2=(N_{\text{c}}^2-1)/(2N_{\text{c}})$, we find that the parabola
depicted in Fig.~1 is lifted and the fixed points vanish for large
gauge coupling. In this case, the system would finally run into the
\xsb\ phase once the gauge coupling has grown large enough. The
question as to whether this mechanism can successfully be applied to a
sector of the standard model is currently under investigation.

\section*{Acknowledgment}
H.G. would like to thank the organizers of this conference
  for creating a lively and stimulating atmosphere. This work has been
  supported by the Deutsche Forschungsgemeinschaft under contract Gi
  328/1-1 and KON 362/2002.

\end{document}